\documentclass{aa}
\usepackage[utf8]{inputenc}
\usepackage[T1]{fontenc}
\usepackage{natbib}
\bibpunct{(}{)}{;}{a}{}{,} 
\usepackage[pdftex=true,plainpages=false,pdfpagelabels,hyperindex=true,colorlinks=true,linkcolor=black]{hyperref}
\usepackage{url}
\usepackage{graphicx}
\usepackage{subfigure} 
\usepackage{xcolor}
\usepackage{geometry}
\usepackage{varioref}
\usepackage[bf]{caption}
\usepackage{longtable}
\usepackage[varg]{txfonts}

\newcommand{\vect}[1]{\ensuremath{\mathbf{#1}}}

\newcommand{\grad}{\ensuremath{\vect{\nabla}}}

\newcommand*{\pd}{\partial}
\newcommand*{\fb}{\frac}

\renewcommand{\Delta}{\triangle}

\renewcommand{\vect}[1]{\ensuremath{\mathbf{#1}}}

\makeatletter
\def\mathcolor#1#{\@mathcolor{#1}}
\def\@mathcolor#1#2#3{%
  \protect\leavevmode
  \begingroup
    \color#1{#2}#3%
  \endgroup
}
\makeatother

\begin{document}

   \title{Tayler-Spruit dynamo simulations for the modeling of radiative stellar layers}

   \author{L. Petitdemange
          \inst{1},
          F. Marcotte \inst{2},
          C. Gissinger \inst{3}
          \and
          F. Daniel\inst{4}
          }

   \institute{LERMA, Observatoire de Paris, PSL Research University, CNRS, Sorbonne Universit\'e, Paris, France.\\
              \email{ludovic.petitdemange@obspm.fr}\\
        \and Universit\'e C\^ote d'Azur, Inria, CNRS, LJAD, France.\\
        \email{florence.marcotte@inria.fr}\\
         \and
             LPENS Laboratoire de Physique de l'\'Ecole Normale Sup\'erieure, ENS, Universit\'e PSL, CNRS, Paris, France.\\
             Institut Universitaire de France\\
             \email{christophe.gissinger@phys.ens.fr}\\
             \and
             LPENS Laboratoire de Physique de l'\'Ecole Normale Sup\'erieure, ENS, Universit\'e PSL, CNRS, Paris, France.\\
              \email{florentin.daniel@phys.ens.fr}}

   \date{Received xxx, accepted xxx}

 
  \abstract
   {Maxwell stresses exerted by dynamo-generated magnetic field have been proposed as an efficient mechanism to transport angular momentum in radiative stellar layers. Numerical simulations are still needed to understand its trigger conditions and the saturation mechanisms.}
   {The present study follows up on \cite{PetitdemangeMG23}, where we reported the first simulations of Tayler-Spruit dynamos. Here we extend the parameter space explored to assess in particular the influence of stratification on the dynamo solutions. We also present numerical verification of theoretical assumptions made in \cite{Spruit02}, which are instrumental in deriving the classical prescription for angular momentum transport implemented in stellar evolution codes.}
   {A simplified radiative layer is modeled numerically by considering the dynamics of a stably-stratified, differentially rotating, magnetized fluid in a spherical shell.}
   {Our simulations display a diversity of magnetic field topologies and amplitudes depending on the flow parameters, including hemispherical solutions. The Tayler-Spruit dynamos reported here are found to satisfy magnetostrophic equilibrium and achieve efficient turbulent transport of angular momentum, following Spruit's heuristic prediction.}
   {}


\titlerunning{Tayler-Spruit dynamos in simulated radiative stellar layers.}
\authorrunning{L.Petitdemange et al.}

   \maketitle
%

\section{Introduction}

Understanding the transport of angular momentum (AM) and chemical elements in stellar interiors is a cornerstone of stellar evolution models \citep{MaederM2000}. In the recent years, asteroseismic data from space-born missions (satellites CoRoT \citep{BaglinABM09} and {\it Kepler}, TESS \citep{Kepler2010,Aguirre2020,RauerCA14}) have provided unprecedented insight into the dynamics of internal layers, by constraining the rotation rates of the envelope and core of thousands of stars \citep{Aerts2019}. In particular, these observations show that considerable spin-down occurs in the radiative cores of evolved stars, through a process that is not explained by current stellar evolution models \cite{ceillierEGM13,christophe18,vanReeth18,EggenbergerDB19,theseLisa}.

While AM transport is expected to be influenced by a diversity of physical mechanisms, including stellar winds, contraction/dilatation, meridional circulation, internal gravity waves and magnetic fields, the latter are particularly suspected to play a major role as their parametrization in stellar codes has been shown to strongly suppress differential rotation \citep{Eggenberger2005}. However, while surface magnetic fields have been detected in the radiative envelopes of many massive and intermediate-mass stars \citep{Wade2016}, magnetic fields in the radiative cores of post-MS stars are still notoriously difficult to constrain despite the promising detection of magnetic signatures on rotational splittings \citep{Prat2020, VanBeeck2020, Mathis2021}. Therefore, the parametrization of magnetically-driven AM transport in 1D stellar codes rely on both theoretical predictions and direct numerical simulations to characterize magnetic fields and the condition for their generation in radiative layers.

Magnetic fields can be generated and sustained against ohmic dissipation in the star internal layers through a dynamo instability, whereby a fraction of the kinetic energy of plasma motions is converted into magnetic energy. In radiative zones however, the mechanisms for dynamo action are poorly understood due to the scarcity of numerical simulations. So far, the only theoretical models for radiative dynamo that have been parametrized in stellar codes are driven by magneto-rotational instability (MRI) and Tayler instability \citep{wheeler2015,griffiths2022}: these parametrizations have shown that,  while MRI is found to strongly enhance chemical mixing, the most efficient at transporting AM is arguably the Tayler-Spruit dynamo mechanism put forward by \cite{Spruit02} (see also \cite{Eggenberger2022}). 

In this model, dynamo action results from a constructive feedback loop between the winding-up of poloidal magnetic fields into toroidal fields by differential rotation ($\Omega$-effect) on the one hand, and a strong, toroidal magnetic field getting unstable to a pinch-type instability \citep{tayler73} on the other hand. In \cite{Spruit02}, a scenario for magnetic field amplification is suggested, where differential rotation generates a strong, axisymmetric toroidal field out of an arbitrarily weak poloidal magnetic field, until the former breaks into non-axisymmetric modes upon reaching a critical amplitude, thus initiating the dynamo loop. This scenario however was never reproduced in numerical simulations despite numerous attempts, so that the possibility for Tayler-Spruit dynamos to operate in radiative stellar layers had remained somewhat controversial \citep{ZahnBM07}. Moreover, even assuming that the classical Tayler-Spruit mechanism is indeed relevant for radiative zones, the rotation rates of subgiant stars was shown to remain unaccounted for \citep{Cantiello2014,Salmon2022}, motivating investigation of alternative, possibly more efficient saturation mechanisms for this dynamo \citep{Fuller19}. In a forthcoming paper \citep{Daniel2023}, we report that the turbulence induced by the dynamo discussed here produces an additional contribution to angular momentum transport via the Reynolds stress tensor. This could significantly affect the profile of rotation, as for some RGB we show that it could be up to one order of magnitude more important than the magnetic torque applied on the flow.

In a recent paper \citep{PetitdemangeMG23}, we reported the first numerical evidence of such dynamo solutions in simulated radiative layers of stars. The initialization of the dynamo loop was found to somewhat differ from the original prediction of \cite{Spruit02}, in that the excitation of the Tayler-Spruit dynamo was nonlinearly achieved through the prior excitation of a weaker, shear-driven dynamo instability. Extensive numerical simulations is still needed to fully understand its trigger conditions and the saturation mechanisms, and the present paper first aims at clarifying the dominant balances at play in the simulated Tayler-Spruit dynamos at steady-state. Second, we extend our exploration of the parameter space and report the existence of various dynamo topologies, depending on the stratification regime, and in particular the corresponding hydrodynamic configuration of the base flow.


\section{Methods}

\begin{figure}
 \includegraphics[width=0.9\linewidth]{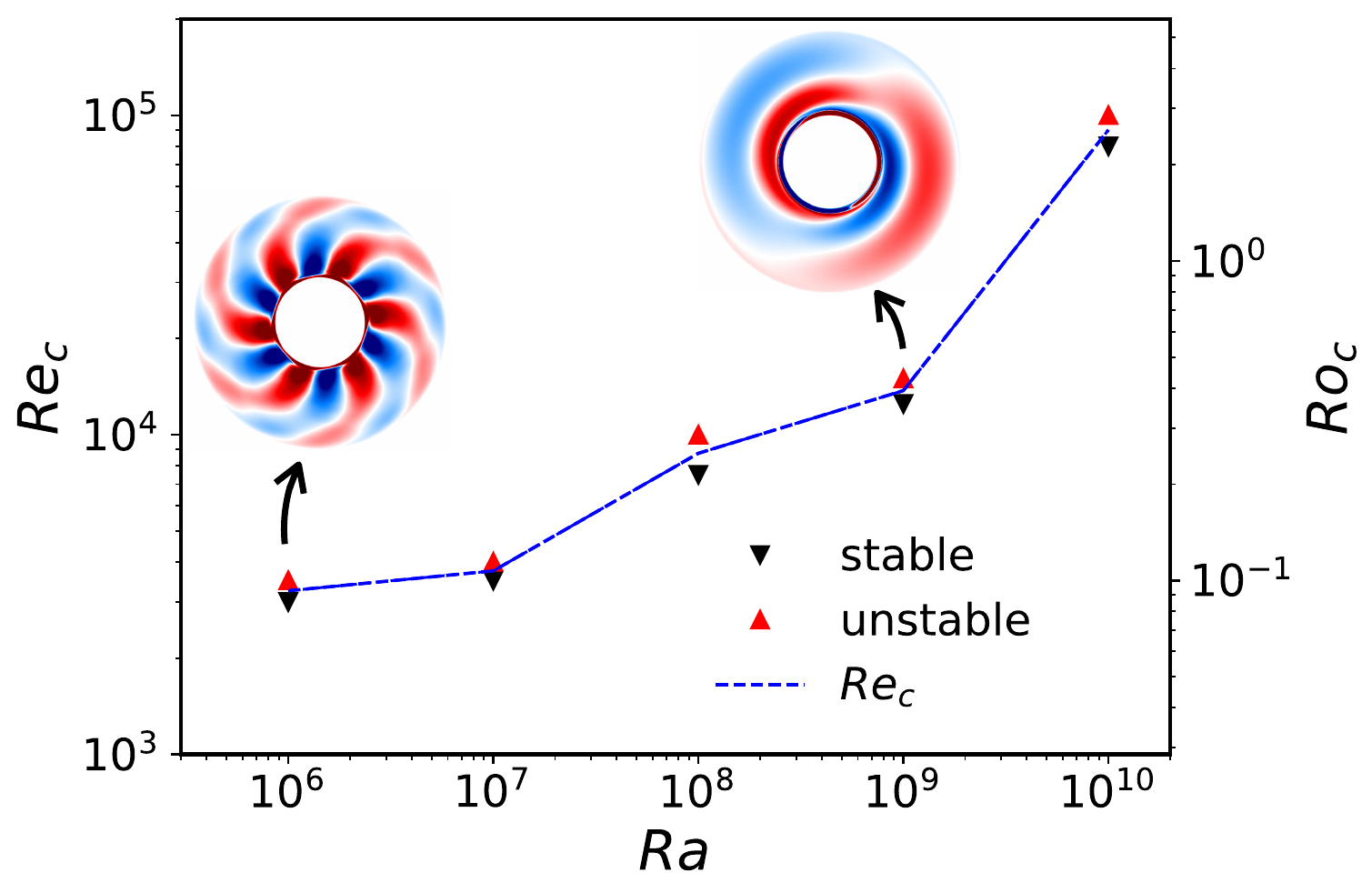}\\
  \includegraphics[width=\linewidth]{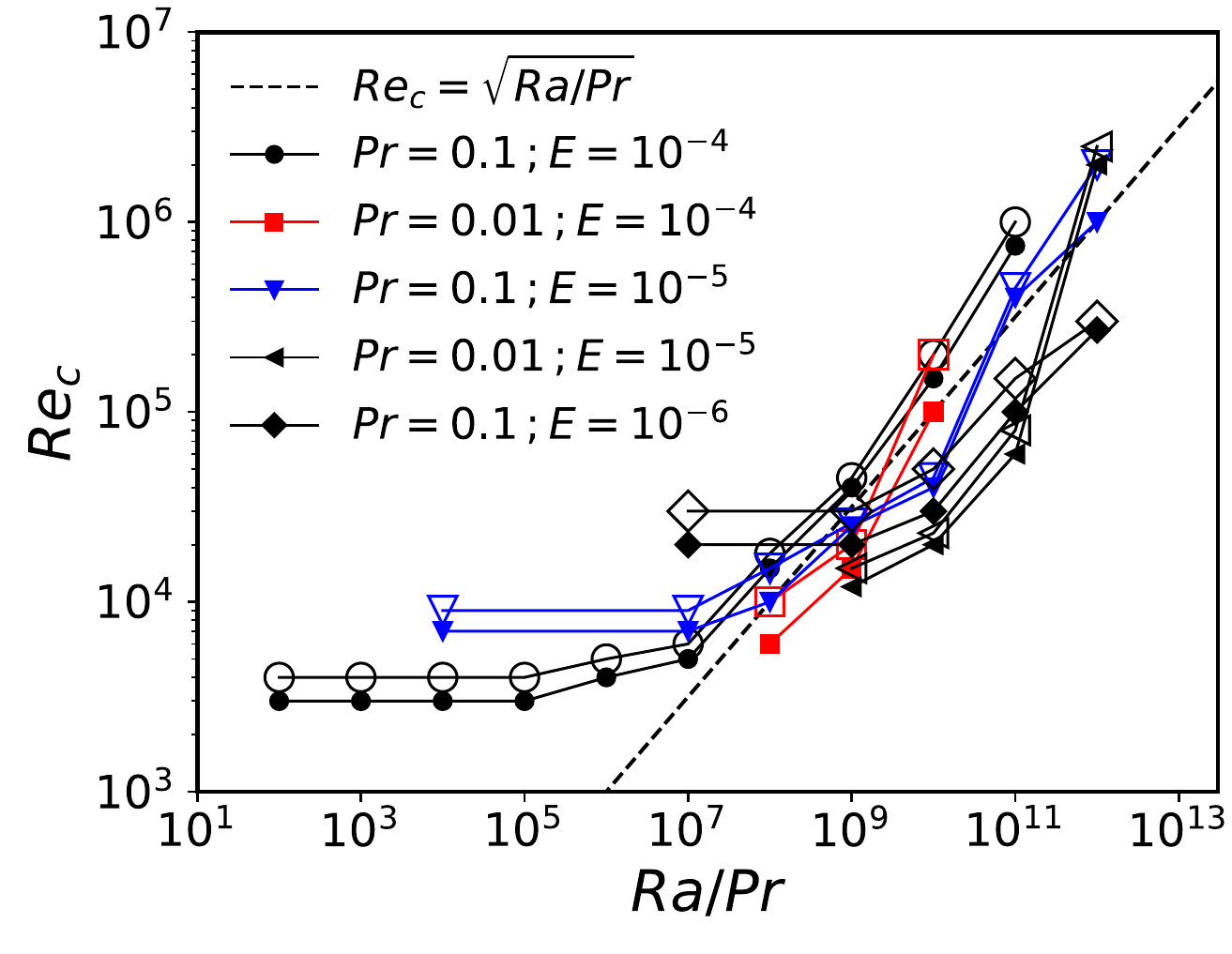}
  \caption{Top: Evolution of the critical Reynolds number $Re_c$ (or equivalently Rossby number $Ro_c$) as a function of stratification intensity (quantified by the Rayleigh number $Ra$). When $Re>Re_c$ (red upwards triangles), non-axisymmetric modes are maintained in time. Conversely when $Re < Re_c$ (dark downwards triangles), the kinetic energies of non-axisymmetric modes exponentially decay with time. The insets highlight the pattern of unstable modes close to the threshold, by showing some colormaps of the radial velocity in the equatorial plane. Simulations parameters: $E=10^{-5}$, $Pr=0.1$.  Bottom: Critical Reynolds number $Re_c$ for the onset of hydrodynamic instability with increasing stratification, for various thermal and rotation parameters.}
\label{StabSCF}
\end{figure}

Our numerical model is presented in detail in the Supplementary material of \cite{PetitdemangeMG23}; we briefly recall it here for clarity. In order to considerably reduce numerical costs, our numerical approach focuses on modeling a star's radiative (stably-stratified) layer, without solving for the complicated couplings with the innermost core regions or the external stellar layers that determine the radial shear profile. Instead, we consider a simple numerical setup where differential rotation in the flow is achieved through entrainment by rigid boundaries: in our model, an electrically conducting fluid with magnetic diffusivity $\eta$ and molecular viscosity $\nu$ fills the gap between two spherical shells rotating about a common axis at different speeds, where $\Omega$ and $\Omega+\Delta \Omega$ are the angular velocities of the outer and inner spheres, respectively. The calculations are performed in the rotating frame of reference where the outer sphere is at rest. {The parameters controlling diffusivities are far from realistic ones due to computational limitations, but we tend to determine systematic behaviours. In addition, it is important to note that the values of effective diffusivities can be increased by several orders of magnitude in strongly turbulent systems such as the tachocline region.}

No-slip boundary conditions are applied on both spheres, along with electrically insulating boundary conditions on the outer sphere, whereas the inner sphere has the same conductivity as the fluid. Stable stratification is achieved inside the fluid by prescribing a positive temperature difference $\Delta T$ between the inner and outer shells, whose temperatures are fixed. The aspect ratio between the inner and outer spheres radii $\chi=r_i/r_o$ is set to $0.35$ throughout the study. We use the Boussinesq approximation to neglect variations in the fluid density except in the buoyancy term, leading to the following MHD equations :
\begin{align}
&\frac{\partial {\bf v}}{\partial t} + ({\bf v} \cdot \grad ) {\bf v}=   \nu \Delta {\bf v} - 2\Omega{\bf e_z}\times{\bf v} - \dfrac{1}{\rho}{\bf \nabla} P +{\bf j}\times{\bf B} + \alpha g_0 r \Theta {\bf e_r}, \\
&\frac{\partial{\bf B}}{\partial t} = \eta\Delta{\bf B} + {\bf \nabla}\times({\bf v}\times{\bf B}), \\ 
&\frac{\partial T}{\partial t} =  -({\bf v}\cdot{\bf \nabla}) T + \kappa\Delta T, \\
&{\bf \nabla}\cdot{\bf v} =0,\\
&{\bf \nabla}\cdot{\bf B}=0
 \label{eq:full_set}
\end{align}
where ${\bf v}$, $P$, ${\bf B}$ and $T$ are respectively the velocity field, pressure field, magnetic field and temperature field. $\Theta$ is the temperature fluctuation accounting for density fluctuations in the buoyancy term, ${\bf e_z}$ the unit vector along the rotation axis and ${\bf e_r}$ the local radial unit vector. The physical properties of the fluid are determined by its magnetic permeability $\mu_0$, its mean density $\rho$, its thermal expansion coefficient $\alpha$ and its thermal diffusivity $\kappa$. $g_0$ denotes the gravitational acceleration at the outer shell. 

In terms of numerical control parameters, the flow regime is entirely described by means of five independent, dimensionless parameters: the Ekman number $E = \nu / \Omega_o r_o^{2}$ quantifying the ratio between viscous force and Coriolis acceleration, the Rayleigh number $Ra = \alpha g_0\Delta T r_o^3 /( \nu\kappa)$ measuring the intensity of thermal forcing (hence the strength of the stratification), the Reynolds number $Re = r_ir_o\triangle\Omega / \nu$ measuring the ratio of inertial to viscous effects, the magnetic Reynolds number $Rm=   r_i r_o\triangle \Omega/\eta$ comparing the respective effects of induction and ohmic dissipation, and the Prandtl number $Pr = \nu / \kappa$ comparing molecular and thermal diffusivities. It is important to emphasize that the extreme flow regimes met in astrophysical flows remain far beyond reach of numerical simulations due to their formidable computational cost. In particular, the Ekman number $E$ is of order $E \sim 10^{-15}$ in stellar interiors, whereas the most intensive high-performance direct numerical simulations (DNS) only reach  $E \gtrsim 10^{-7} - 10^{-8}$ so far \citep{schaefferJNF17}. Simulations in the strongly stratified regime also require a very high spatial resolution as stratification decreases the size of flow structure in the radial direction, and the timestep must be decreased in order to take into account the possibility of internal gravity waves. Nevertheless, 3D DNS can now achieve fully turbulent, strongly stratified flow regimes due to the ever-increasing availability of numerical ressources. Moreover, our simulations systematically span a large range of parameters in order to determine useful scaling laws. 

The numerical resolution is chosen large enough so that physical viscosity always dominates over grid viscosity in our non-ideal simulations, our results are therefore independent of the resolution. All the simulations reported in this study were carried out using the {\sc Parody-JA} code \citep{Dormy1998,Aubert2008} coupled with the ShTns library \citep{Schaeffer2013}. {\sc Parody-JA} uses finite-difference discretization in the radial direction and spherical harmonics decomposition. The number of radial gridpoints (${n_r}$) used in the fluid domain is $288<n_r<360$, and the maximal degree ($l_\text{max}$) and order ($m_\text{max}$) of the spherical harmonics decomposition are $128<l_\text{max}<198$ and $58<m_\text{max}<128$, respectively.

\section{Comment on hydrodynamic states}

Before turning to the dynamo problem, it is interesting to briefly consider the geometry of the (purely hydrodynamic) base flow depending on the rotation and stratification parameters. 

Weakly-stratified, rapidly-rotating flows are dominated by a force balance between Coriolis acceleration and pressure gradient, a situation referred to as the geostrophic equilibrium. This dominant balance results in the flow being (at lowest order) invariant along cylinders coaxial with the rotation axis \citep{Proudman1956}. In spherical Couette flow (and below the shear instability threshold), the fluid located outside the notional tangent cylinder encompassing the inner sphere at the equator is co-rotating with the outer sphere. Inside the tangent cylinder (and far from both the boundaries), the fluid rotates at rate $\Omega+\Delta \Omega/2$. The velocity jump across the tangent cylinder is accomodated through a series of nested, free shear layers (the Stewartson layers \citep{stewartson1966}), whose thicknesses scale like powers of the rotation parameter $E$.  (\cite{marcotteDS16} have provided a complete description of this problem.) When the shear parameter becomes sufficiently large, destabilization of a Stewartson shear layer results in breaking the flow axisymmetry, while invariance along the rotation axis remains largely preserved.

Strongly-stratified, slowly-rotating flows on the other hand exhibit a spherical geometry due to leading effect of buoyancy: this corresponds to the typical regime where the \textit{shellular} approximation is meaningful, or in other words, that the angular velocity can be considered (at leading order) invariant in the horizontal direction \citep{zahn92}. The transition in flow geometry depends on the relative importance of stratification and global rotation: as shown by \cite{philidet2019} though, the critical parameter is not the frequency ratio $N/\Omega$ between the buoyancy frequency $N=(\alpha g\Delta T/(r_o-r_i))^{1/2}$ and the global rotation rate $\Omega$, but the quantity $ Q\equiv Pr\left(\frac{N}{\Omega}\right)^2$ - equivalently expressed here as $Q\equiv E^2Ra/(1-\chi)$. This numerical result of \cite{philidet2019} can be retrieved with a simple dimensional analysis of the problem. Balancing Coriolis acceleration and buoyancy in the equatorial plane yields
  \begin{equation}
2\Omega u_\phi \sim \alpha g\Theta,
\label{eq:NSbuoyancy}
  \end{equation}
  where $u_\phi$ and $\Theta$ are typical values for the azimuthal velocity and the thermal perturbation, respectively. In the Boussinesq approximation, we can estimate the thermal perturbation at steady state as
\begin{equation}
u_r\frac{dT_s}{dr}\sim \kappa\bigtriangleup \Theta,
\label{Teq}
\end{equation}
where $u_r$ is a typical radial velocity; furthermore, the azimuthal component of (1) in the equatorial plane provides
\begin{equation}
2\Omega u_r\sim \nu\bigtriangleup u_\phi.
\label{NSphieq}
\end{equation}
Combining \ref{eq:NSbuoyancy},\ref{Teq} and \ref{NSphieq} yields 
\begin{equation}
Q \equiv \frac{\alpha g}{\Omega^2} \frac{\nu}{\kappa} \frac{dT_s}{dr} \sim 4.
\end{equation}
This qualitative argument further suggests that the crossover regime corresponds to $Q$ of order unity, which happens to be relevant for example in the case of the solar radiative zone due to the smallness of the Prandtl number $Pr$ (with $Q_{\text{Sun}} \sim 1.3$).

Fig. \ref{StabSCF} shows that, increasing the stratification (controled by the Rayleigh number $Ra$) when the rotation parameter $E$ and the thermal Prandtl number $Pr$ are fixed tends to stabilize the non-axisymmetric shear instability, which also develops at decreasing wavenumbers. Unsurprisingly, the stability threshold $Re_c$ scales as $Re_c \sim \sqrt{Ra/Pr}$, which is equivalent to a Richardson number $Ri$ of order unity \citep{Richardson1920}:
\begin{equation}
Ri \equiv \fb{N^2}{(\Delta \Omega)^2}=\fb{Ra}{PrRe^2} \sim 1.
\end{equation}

\section{Stratification, rotation and dynamo morphology: a route to Tayler-Spruit dynamos}

\begin{figure}
    \includegraphics[width=\linewidth]{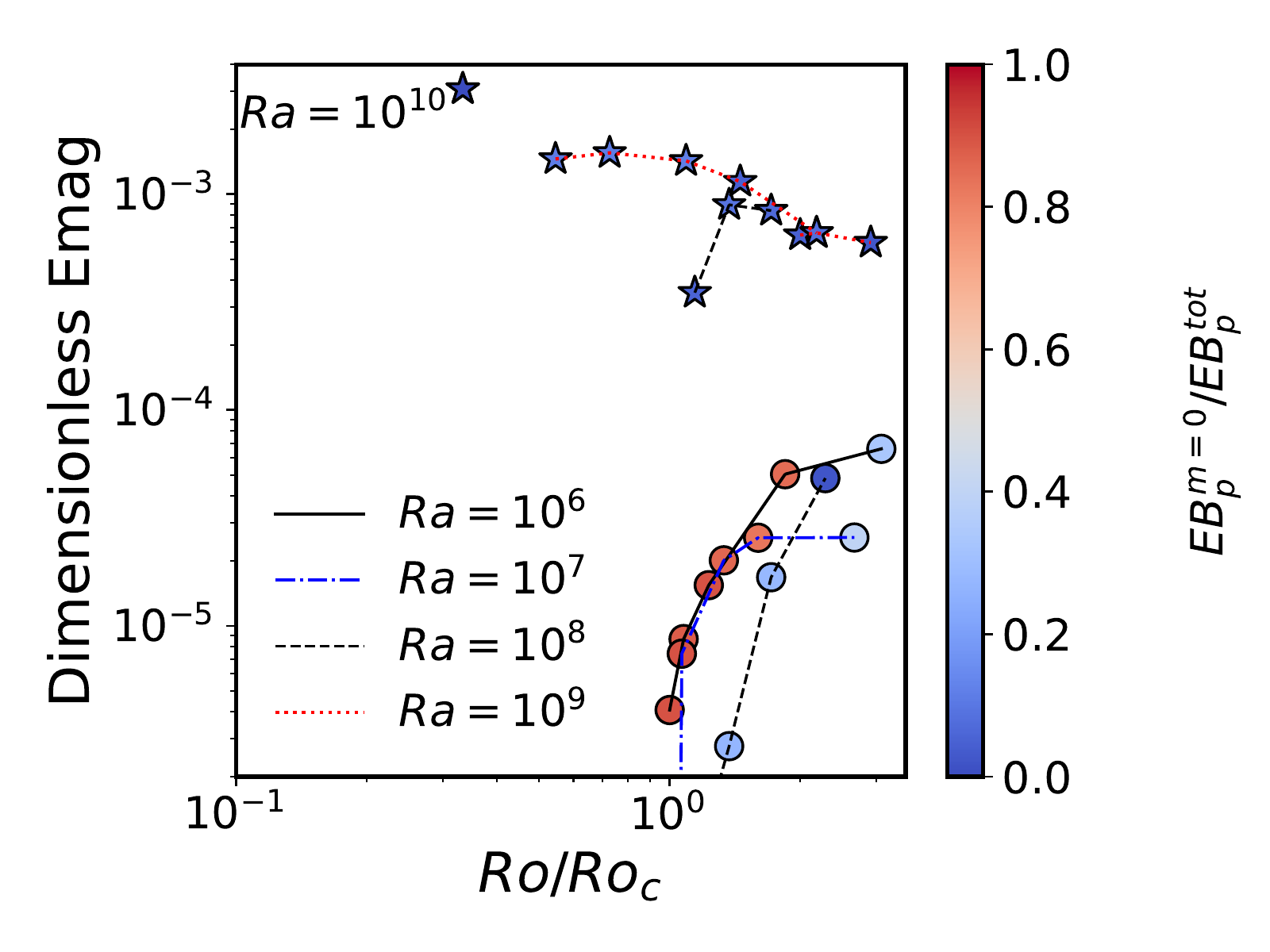} \\
  \caption{Time averaged magnetic energy  for dynamo simulations performed with different parameters and initial conditions for the magnetic field. Star symbols correspond to models in which the axisymmetric toroidal magnetic component exceeds 80\% of the total magnetic energy (toroidal dynamos). Fixed simulations parameters: $E=10^{-5}$, $Pr=0.1$, $Pm=1$.}
  \label{Re_mag}
  \end{figure}

\begin{figure}
      \begin{center}
  \includegraphics[width=\linewidth]{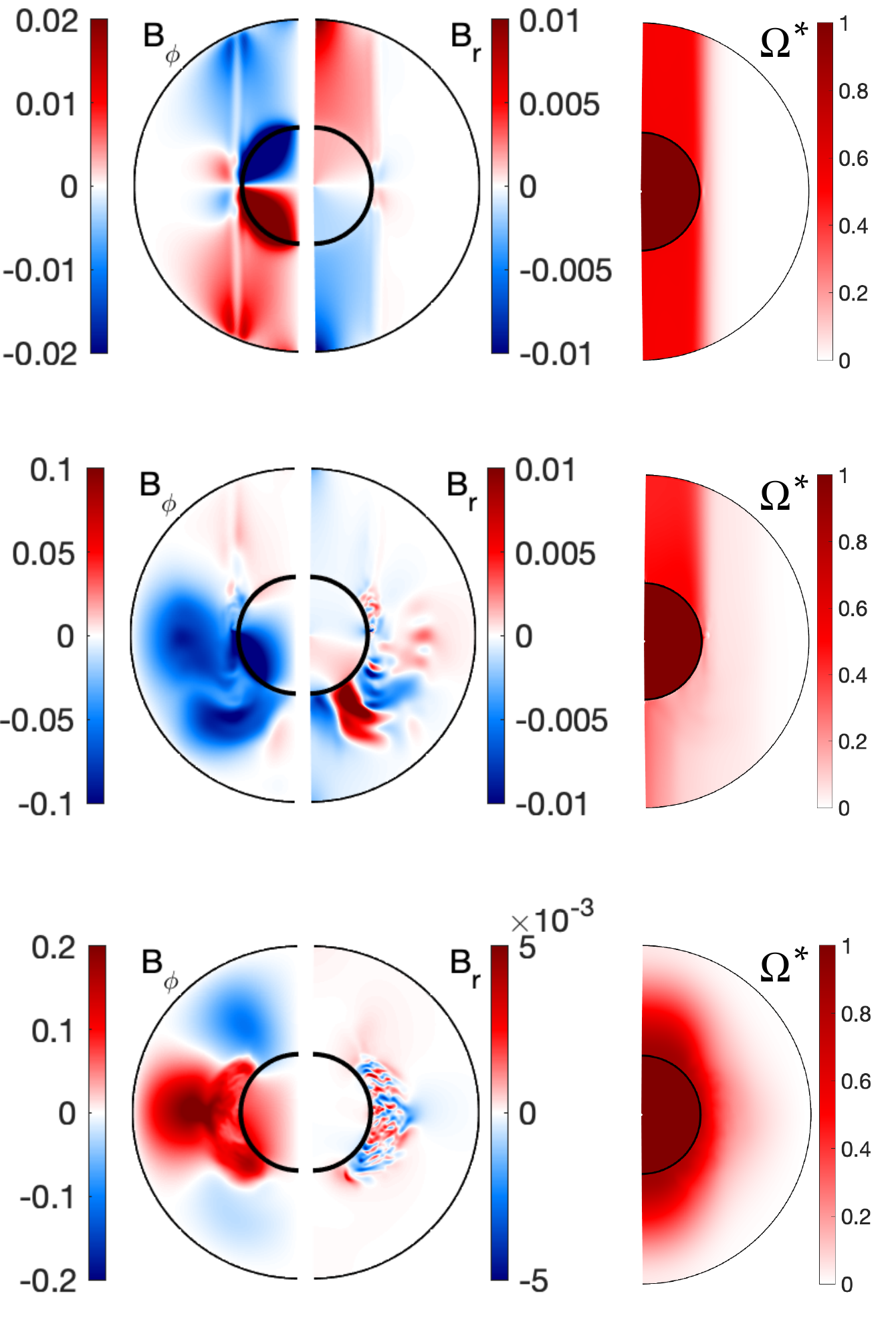}
    \caption{Meridional sections of the axisymmetric components of azimuthal magnetic $B_\phi$, radial magnetic field $B_r$, and angular velocity $\Omega^\star$ (normalized by $\Delta \Omega$). \textit{Top}: Weakly-stratified, dipolar dynamo at $E=10^{-5}$, $Ra=10^7$, $Re=5000$, $Pr=0.1$, $Pm=1$. \textit{Middle:} Hemispherical  dynamo at $E=10^{-5}$, $Ra=10^8$, $Re=12000$, $Pr=0.1$, $Pm=1$. \textit{Bottom:} Strong toroidal (Tayler-Spruit) dynamo at $E=10^{-5}$, $Ra=10^{10}$, $Re=3.10^4$, $Pr=0.1$ and $Pm=1$.}
    \label{fig_maps}
    \end{center}
    \end{figure}

The hydrodynamic base states described above are found to power a diversity of self-sustained radiative dynamos, depending on the ratio stratification-to-rotation on the one hand, on the initial condition for the magnetic field on the other hand. The variety (both in amplitude and in structure) of the dynamo-generated magnetic fields  is exemplified by the diagram in Fig.~\ref{Re_mag}, which summarizes the study of the dynamo bifurcation for fixed diffusivity ratios $Pr=0.1$,$Pm=1$ and rotation parameter $E=10^{-5}$, when the Rayleigh number controling the degree of stratification is varied over 4 order of magnitudes. The dynamo morphology can be mainly classified into three categories, illustrated in Fig.~\ref{fig_maps}: toroidal, dipolar or hemispheric dynamos. Toroidal dynamos are characterized here by a strongly dominant axisymmetric toroidal component (representing more than 80 per cent of the magnetic energy), {and a largely subdominant, non-axisymmetric poloidal component dominated by the azimuthal mode $m=1$. These dynamos correspond to Tayler-Spruit dynamos \citep{Spruit02}, numerically reproduced for the first time in \citep{PetitdemangeMG23} and presented in detail therein. Dipolar dynamos on the other hand refer to dynamos where the poloidal energy only slightly dominates over the toroidal energy,} but the former is dominated by the axial dipole. Finally, hemispherical dynamos are characterised by a strong asymmetry between the two hemispheres.

The dipolar dynamos are observed at sufficiently low $Ra$, where stratification has a minor influence on the MHD flow; these dynamos are similar to those reported in \citet{guervillyC10} without stratification. In this low-$Ra$ regime, neither the magnetic field morphology nor its amplitude are found to depend on the initial conditions. The relative weakness of the dynamo-generated magnetic fields results in the flow not being significantly affected by the dynamo: although the total kinetic energy is slightly reduced, the dominant non-axisymmetric component of the velocity field remains unchanged compared to the purely hydrodynamic simulations.

Stratification effects become non-negligible as $Ra$ increases to $Ra=10^8$ in Fig.~\ref{Re_mag}: two distinct branches of dynamo solutions can be observed depending on the magnetic seed field prescribed initially. Specifically, tiny, random magnetic fields are amplified and evolve toward saturated dipolar dynamos (dashed line with circle symbols in Fig.~\ref{Re_mag}), whereas systems initialized with a stronger, well-chosen magnetic field spontaneously evolve toward a new equilibrium, corresponding to a toroidal (Tayler-Spruit) dynamo (dashed line with star symbols). These two branches appear as soon as $Re/Re_c$ becomes slightly greater than one, meaning that both solutions require the hydrodynamic instability to set in for a magnetic field to be maintained by dynamo action. For $Re<Re_c$, the magnetic energy and the kinetic energy of all non-axisymmetric modes eventually undergo exponential decay whatever the tested initial condition. Moreover, the amplitude of the saturated magnetic field is more than one order of magnitude larger for Tayler-Spruit dynamos than for dipolar dynamos (and almost three orders of magnitude for $Re/Re_c < 1.5$). As a consequence, Tayler-Spruit dynamos are observed to trigger magnetohydrodynamic turbulence whereas dipolar solutions are not. As shown in \citep{PetitdemangeMG23}, the associated transport of angular momentum tends to suppress the shear in the bulk of the flow, where magnetic activity is most intense.

The rich dynamo topology of this moderately stratified regime is further illustrated by the emergence of hemispherical dynamo solutions. Fig.~\ref{fig_maps} shows meridional maps of the angular velocity and magnetic fields: in such dynamo solutions, magnetic activity is nearly absent from one hemisphere and the flow corresponds to that of the hydrodynamic state, with columnar vortices aligned with the tangent cylinder. In the other hemisphere, a strong, large-scale axisymmetric toroidal magnetic field builds up at mid-latitudes. In comparison, the radial component has a smaller typical length scale and it is dominated by its non-axisymmetric components. The field strength for these dynamos is strong enough to considerably modify the flow structure in the hemisphere influenced by the magnetic field: the cylindrical symmetry of the flow is broken and the non-axisymmetric components of the radial velocity exhibit the same spatial distribution as the magnetic field.

As $Ra$ further increases though, a crucial feature of Tayler-Spruit dynamos is that (some!) strong magnetic fields can be maintained, not only below the linear instability threshold for dynamo action, but even below the shear instability threshold \citep{PetitdemangeMG23}. In the $Ra=10^9$ series presented in Fig.~\ref{Re_mag} for example, no initially weak magnetic fields are amplified up to $Re/Re_c \sim 1.75$. Above this threshold, initially weak seed fields are amplified by dynamo action and develop into strong, toroidal fields, which in turn promote turbulent fluid motions. (Note that all the series shown in this particular figure correspond to $Pm=1$. Another type of dynamo, morphologically similar to the Tayler-Spruit dynamos presented here, but considerably weaker and essentially laminar, can be obtained at lower $Pm$ and was reported in \citep{PetitdemangeMG23}.) These strong dynamo-generated fields could be maintained down to $Re \sim 0.5 Re_c$, modifying the flow so as to sustain the turbulent motions that power them even below the (hydrodynamic) instability threshold. In \cite{Daniel2023}, we show by a $Pm$ study that the relevant criterion controlling how low the system can go before losing dynamo action (in terms of differential rotation) is actually a constant magnetic Reynolds number, whose value is fixed by the global rotation and stratification, in very good agreement with what was proposed by \cite{Spruit02}. It is therefore likely that, in a real star, as the $Rm$ number is large, Tayler-Spruit dynamos could flatten rotation profiles across the radiative zone and still operate even when the shear becomes comparatively weak.

This subcriticality of the Tayler-Spruit mechanism was certainly instrumental to numerically reproduce these particular dynamo solutions, and may explain why they have long eluded numerical investigation. Indeed, Tayler-Spruit dynamos here can be obtained out of initially weak, random seed fields only when the shear is sufficiently strong and the base flow axisymmetry broken by hydrodynamic instability, meaning that $Re \gg Re_c \sim \sqrt{Ra/Pr} = N/(\Omega E)$. While this is never a restrictive condition in a (real) stellar interior, where the expected Reynolds numbers are of order $Re \sim 10^{10}$ or larger, it certainly is for numerical simulations, for which the computational cost associated with the shear-unstable regime rapidly becomes overwhelming at large $Ra$ (strong stratification). In other words, attempting to grow a numerical Tayler-Spruit dynamo directly out of weak, random magnetic fields using, for example, a solar-like ratio $N/\Omega \sim 100$ and a reasonably small rotation parameter $E\ll 1$ is practically hopeless. This may seem particularly surprising since Tayler-Spruit's theory predicts dynamo action out of vanishingly small initial perturbations, due to prior amplification through $\Omega$-effect: while this scenario may apply in true radiative stellar layers, the required flow regimes might not allow to see such a scenario in numerical simulations.

\begin{figure}
\includegraphics[width=\linewidth]{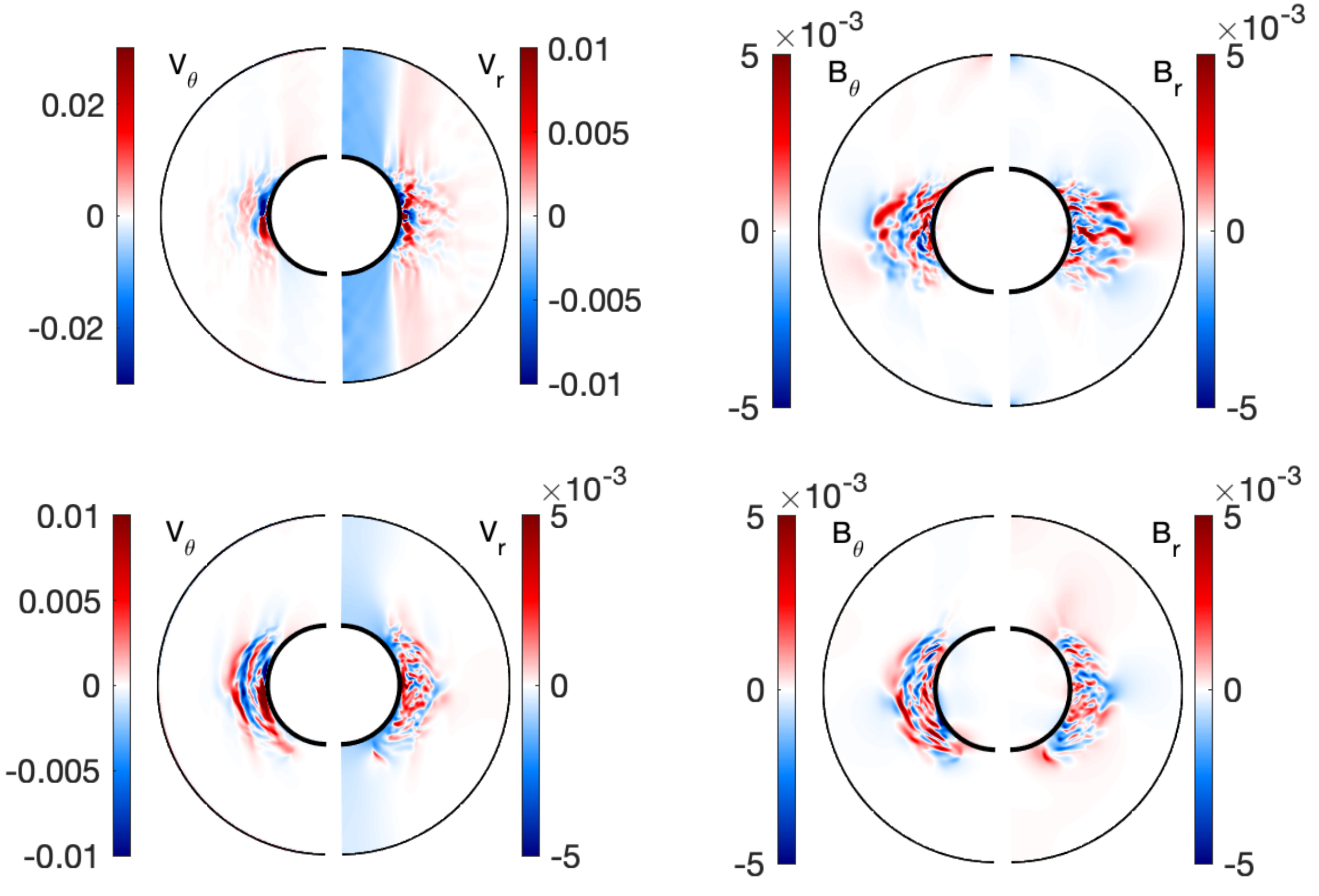} 
\caption{Meridional sections of axisymmetric latitudinal and radial components of the velocity field (on the left) and magnetic field (on the right). Parameters:  $Re=2.75\times 10^4$ and $Ra=10^9$ (uppel panel; $Q=0.15$);  $Re=3.10^4$ and $Ra=10^{10}$ (lower panel; $Q=1.5$), other control parameters are fixed as per Fig.\ref{fig_maps}.}
\label{cyl2sph}
\end{figure}

In our approach instead, we exploit the subcriticality of Tayler-Spruit dynamos to reach large $N/\Omega$ regimes by first addressing the computationally more accessible supercritical flow regime ($Re > Re_c$) for $N/\Omega=O(1)$. Once a dynamo builds up, the resulting magnetic and velocity fields at steady state are used to initialize a new simulation, where the control parameters defining the flow regime are slightly modified. Once the fields have adjusted to the new flow conditions and settled into a steady state, the latter is used to initialize a new simulation where the control parameters are further modified, and so on.  Following the dynamo manifold requires that the control parameters vary slowly while exploring the parameter space, due to the subcritical nature of the dynamo. Despite the numerical constraints, we find that Tayler-Spruit dynamos are surprisingly robust and actually span a wide spectrum of flow regimes. While many simulations here are performed with small ratios $N/\Omega$ that are relevant for rapid rotators, it was possible to obtain a few simulations with $N/\Omega$ up to 20 and 50, thus approaching the solar ratio ($N/\Omega|_{\text{Sun}}\sim 100$). These simulations correspond to $Q=62$ and $Q=246$, respectively, meaning that the base flow has very clear spherical symmetry. The velocity and magnetic maps shown in Fig. \ref{cyl2sph}, both corresponding to Tayler-Spruit simulations, exemplify the clear emergence of this symmetry as $Q$ increases and the robustness of the dynamo mechanism with respect to the flow geometry.

\begin{figure}
\begin{tabular}{cc}
\includegraphics[width=0.9\linewidth]{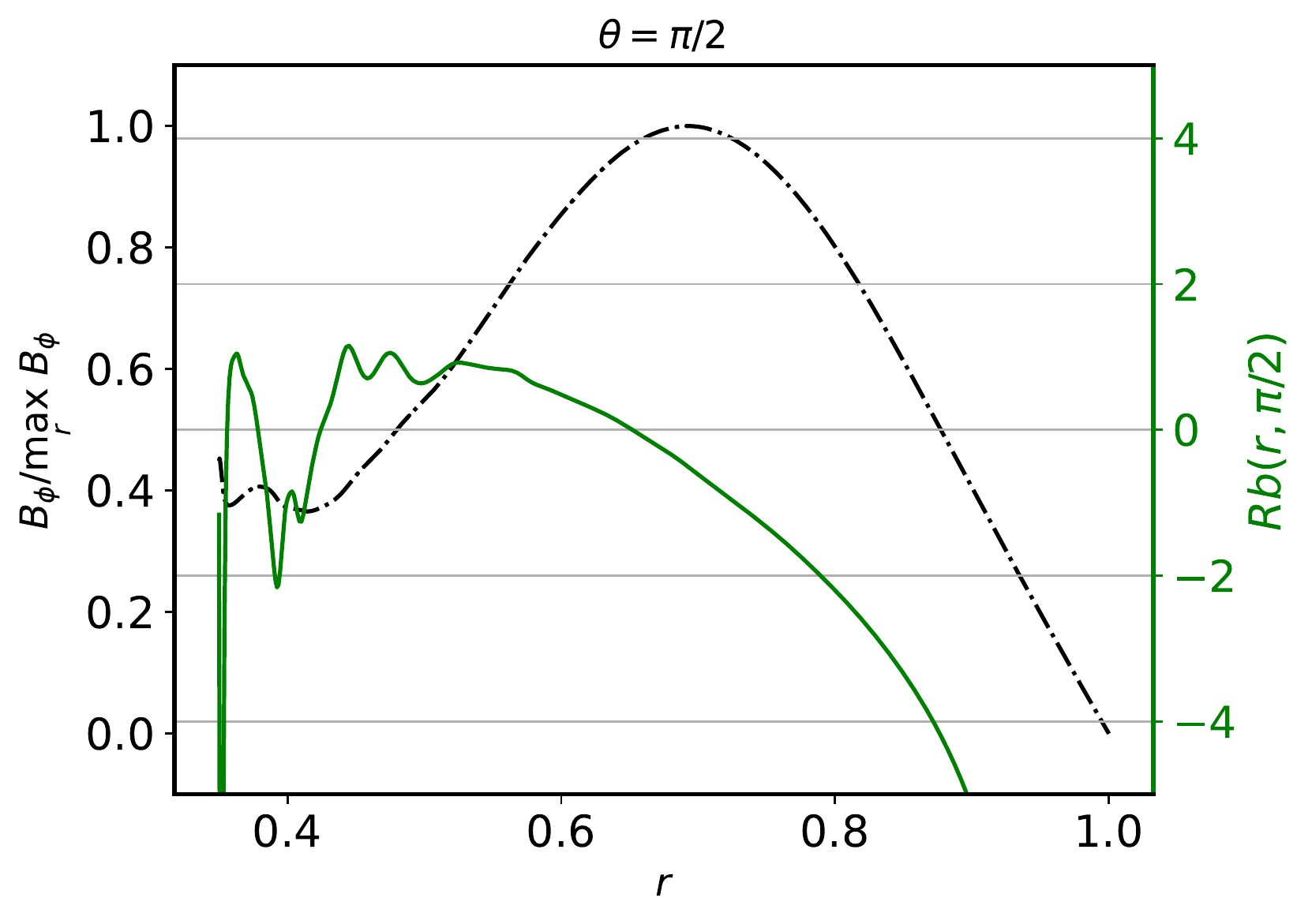} \\
\includegraphics[width=0.5\linewidth]{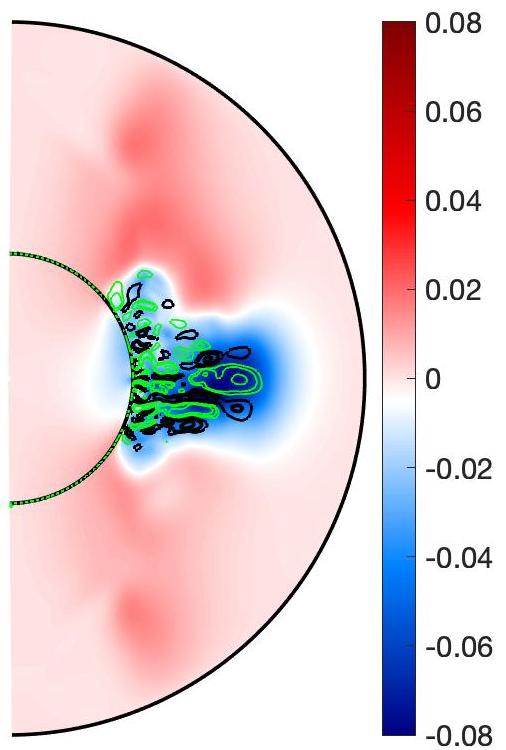}
\end{tabular}
\caption{{Top: Radial profile of the magnetic Rossby number Rb and the axisymmetric azimuthal field $\bar{B_\phi}$. Bottom: isovalues of the non-axisymmetric component $B_r^n$ (positive in green and negative in black) are superposed to a  meridional section of axisymmetric toroidal component of the magnetic field at a given time during the saturation phase of the dynamo. Parameters:  $Re=2.75\times 10^4$ and $Ra=10^9$, other control parameters are fixed as per Fig.\ref{fig_maps}. Note that these parameters correspond to the fiducial run shown in Fig. 2 of \citet{PetitdemangeMG23}.}}
\label{fig_profile_Rb}
\end{figure}

{It is important to note that, while our simulations present a few typical properties indicating that a Tayler instability is at work to generate the non-axisymmetric fluctuations whose nonlinear interactions close the dynamo loop, the possibility for the azimuthal field to become destabilized through azimuthal magnetorotational instability (AMRI) in some parameter regimes and thus feed the dynamo cannot be ruled out. For example, simulations by \citet{Guseva17} have shown that AMRI can drive a dynamo and sustain MHD turbulence in a cylindrical shear flow. Motivated by the design of experiments, the stability of an azimuthal, background magnetic fields has been investigated in cylindrical geometry, considering different radial variations of the background magnetic and velocity field \citep{kirillovSF14,Rudiger18}. These studies have shown that purely azimuthal fields can not only trigger both Tayler instability and AMRI, but that their instability domains actually overlap in some regions of the parameter space, making the distinction between both mechanisms somewhat difficult. In the present case, several indicators point toward the Tayler instability: as we have shown in \citet{PetitdemangeMG23}(Fig. 2), the secondary energy growth characterizing the onset of strong, toroidal dynamo action occurs at the time where the local stability criterion for Tayler instability is met, as quantified using the local Elsasser number. Secondly, at high $Pm$, \citet{Daniel2023} find the minimum shear rate (controlled by the Rossby number) that is required to maintain these strong dynamo solutions to be in good agreement with the predictive scaling of \cite{Spruit02}. Finally, the radial profile of the corresponding axisymmetric, azimuthal magnetic field is shown in Fig.\ref{fig_profile_Rb}(Top), where the so-called magnetic Rossby number $Rb \equiv r^2 \partial_r(\bar{B_\phi}/r)/(2\bar{B_\phi})$ measures the local steepness of the axisymmetric component of the azimuthal field $\bar{B_\phi}$: comparison with the meridional map of non-axisymmetric magnetic fluctuations in Fig.\ref{fig_profile_Rb}(Bottom) reveals that the latter develop only when the large-scale field increases sufficiently fast in radius, as in Spruit's scenario. More specifically, non-axisymmetric fluctuations take place in the region, relatively close to the inner sphere, where $Rb > 0$: this is always found to promote Tayler instability in the setups considered by \citet{kirillovSF14}. Conversely, the flow domain where $Rb < -0.5$, which is \textit{a priori} expected to be prone only to AMRI following \citet{kirillovSF14}, remains well off the dynamo active region. Some caution is needed when interpreting the present simulations in the light of these studies, which consider unstratified fluids and a different geometry, and it would be desirable to investigate how these results pertain to the stratified case and spherical geometry in a future work.} 

 \section{Saturation of the Tayler-Spruit dynamo and angular momentum transport}

A major consequence for the possible existence of Tayler-Spruit dynamos in radiative stellar layers is the resulting enhancement of angular momentum (AM) transport. Quantitative prediction of AM transport however requires understanding the saturation processes at play in the Tayler-Spruit dynamo loop. In the seminal paper \cite{Spruit02}, the author derives a theoretical prediction for quantifying the azimuthal Maxwell stress $T \sim B_r B_\phi /\mu$ using dimensional analysis. The derivation can be summarized as follows: first, the azimuthal component of the induction equation at steady state provides a balance between amplification of the azimuthal field by the $\Omega$-effect on the one hand, and its (effective) damping through Ohmic diffusion on the other hand: at leading order,
\begin{equation}
r (B_r \cdot \nabla) \Omega \sim \fb{\eta_{\text{eff}}}{l_r^2}B_\varphi,
\end{equation}
where $l_r$ is the typical radial (vertical) lengthscale for magnetic activity, $B_r,B_\phi$ the typical amplitudes of the radial and azimuthal dynamo-generated field, and $\eta_{\text{eff}}$ the effective (i.e. turbulent) magnetic diffusivity. Simple dimensional arguments stemming from mixing length theory suggest that $\eta_{\text{eff}} \sim \sigma l_r^2$, where $\sigma=B_\phi^2/\rho\mu\Omega r^2$ is the growth rate of Tayler instability \cite{Spruit02}, so that
\begin{equation}
B_r q \Omega \sim \fb{B_\phi^3}{\Omega \rho \mu r^2},
\label{ind}
\end{equation}
where we follow Spruit's notation for the dimensionless shear rate $q \equiv r \pd_r \Omega/\Omega$. The typical radial scale for dynamo activity can be chosen as the largest unstable length scale with respect to Tayler instability,
\begin{equation}
l_r \sim \fb{B_\phi}{\sqrt{\rho\mu}N}.
\label{lr}
\end{equation}
(The scaling above derives from the consideration that stratification suppresses instability at the largest scales, and is obtained for the case $\kappa=0$. A classical way of reinstating thermal diffusivity when $\kappa \gg \eta \neq 0$ is to observe that, since thermal diffusion acts on a faster timescale than the dynamo, its effect is merely to partly suppress temperature gradients, thus decreasing the effective buoyancy frequency $N~\leftarrow~N_e=N\sqrt{\eta/\kappa}<N$ \cite{zahn74}.)

\begin{figure}
  \includegraphics[width=0.8\linewidth]{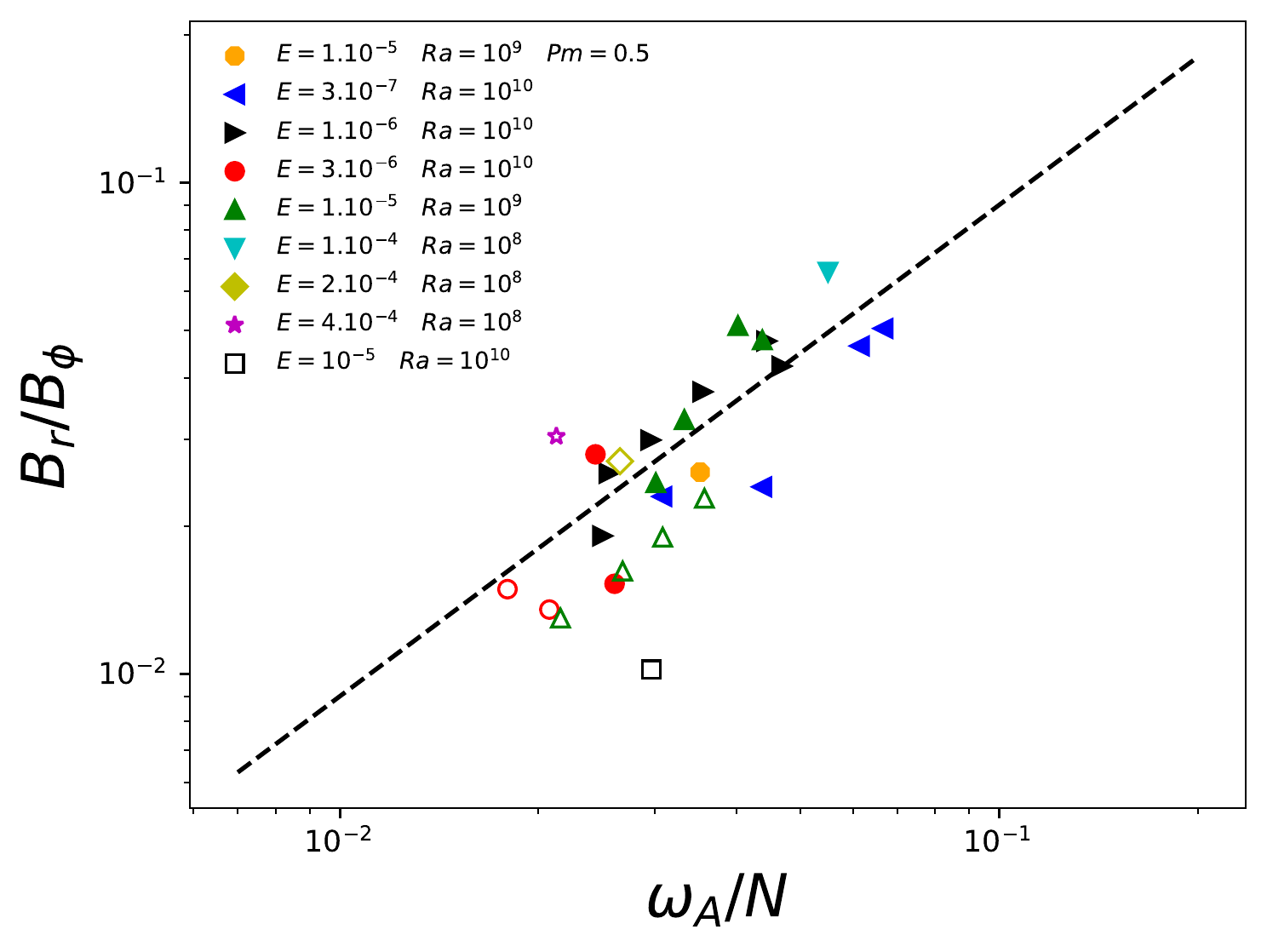}\\
  \includegraphics[width=0.9\linewidth]{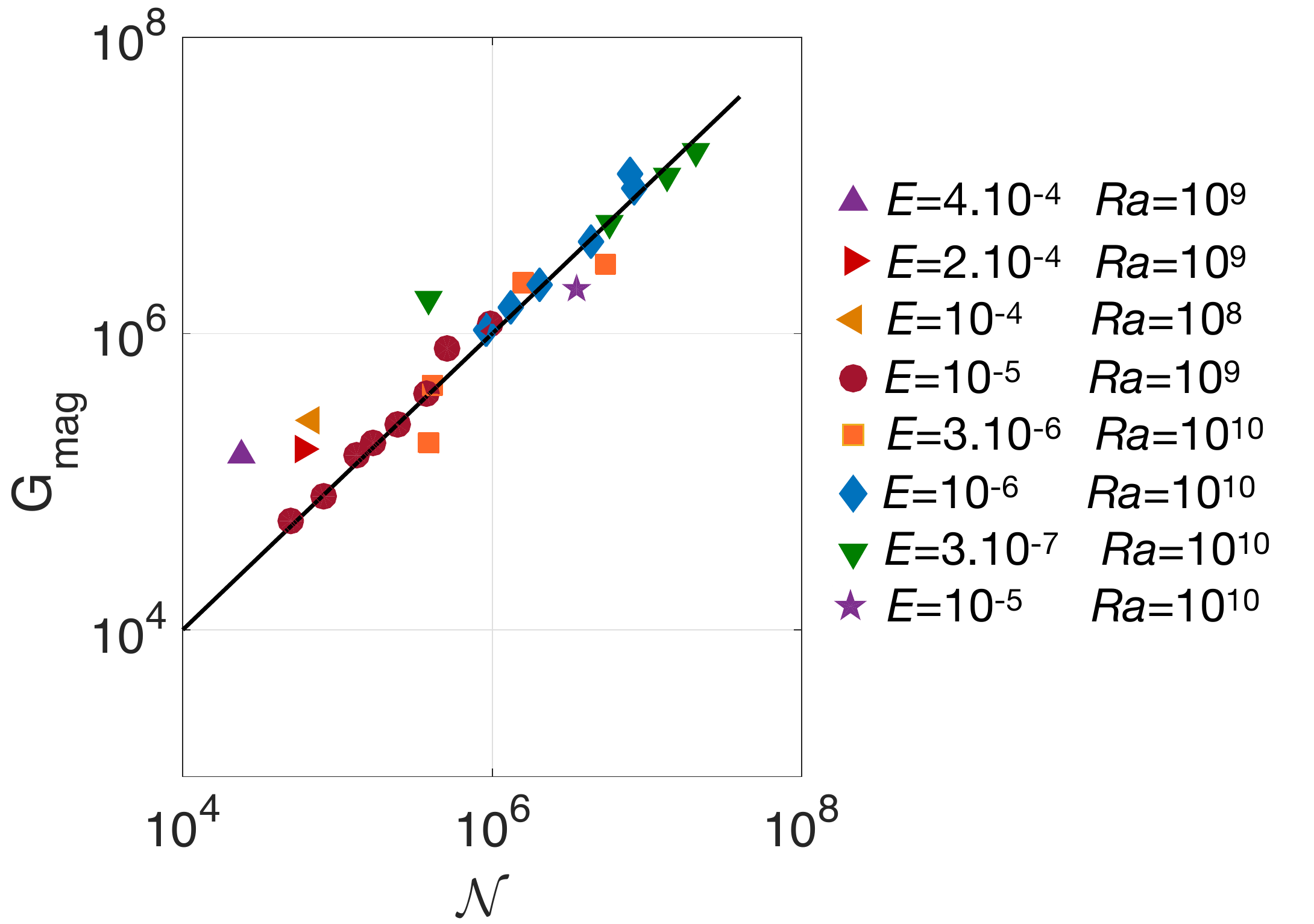}
  \caption{Top: Dimensionless ratios $B_r/B_\phi$ versus $\beta B_\phi/\sqrt{\rho\mu}rN$, for the dynamo simulations shown in \cite{PetitdemangeMG23} and for our additional, highly-stratified model. $B_r$ and $B_\phi$ here are determined by averaging the energy of the $m=0$ poloidal and toroidal magnetic fields over space and time (at steady state). Here $\beta=1.8$ is a fitting coefficient. Note that empty symbols denote simulations in only weakly supercritical regime (as quantified arbitrarily here by $Re/Re_c \le 5/3$): because the flow regime remains close to the onset of instability, the relevance of diffusionless scaling is subject to caution. Bottom: Scaling of the dimensionless magnetic torque with Spruit's diffusionless prediction Eq.\ref{Gmag} (shown with the black line). The data include the simulations shown in \citep{PetitdemangeMG23} and an additional, highly-stratified model with $Ra~=~10^{10}$. }
\label{fig_divB}
\end{figure}

A last equation is needed to close the system and evaluate $T$. For this, \cite{Spruit02} considers the rate at which radial magnetic field is generated from the axisymmetric, azimuthal field by Tayler-unstable displacements to write:
\begin{equation}
\fb{B_r}{l_r} \sim \fb{B_\phi}{r}, 
\label{dB}
\end{equation}
(This can be obtained by noting that
\begin{equation}
\fb{\partial B_r}{\partial t} \sim \fb{B_\phi}{r} \fb{\partial u_r}{\partial \phi} \implies B_r \sim (\tau u_r) \fb{B_\phi}{r} \sim l_r \fb{B_\phi}{r},
\label{dB2}
\end{equation}
where $\tau$ is the production timescale of $B_r$ through Tayler instability, and $u_r$ the radial velocity associated with unstable displacements.\footnote{Note that the published version of the present article (to appear soon) contains a mistake in the RHS of Eq. 15 (first equation), which has been corrected here.})
Combined together Eqs.\ref{ind}, \ref{lr} and \ref{dB} yield the predictive scaling law for the Maxwell stress in the diffusionless case:
\begin{equation}
T = \fb{B_r B_\phi}{\mu}\sim\rho\Omega^2r^2q^3\left(\fb{\Omega}{N}\right)^4.
\label{scaling}
\end{equation}
The arguments used to derive Eq.~\ref{scaling} in \cite{Spruit02} have been largely debated (e.g. \cite{ZahnBM07}), owing to the fact that the most unstable azimuthal wavenumber with respect to Tayler instability is not $m=0$ but $m=1$. As a result, the radial field $B_r$ produced in Eq.\ref{dB2} should be dominated by the $m=1$ component, which can hardly (directly) replenish the axisymmetric $B_\phi$ through $\Omega$-effect in Eq.\ref{ind}. This is no theoretical obstacle in a turbulent flow though, since non-linear interactions between small-scale velocity and magnetic fluctuations can produce the required axisymmetric fields through mean-field effect \citep{moffatt}. The scaling Eq.~\ref{dB} could be also recovered using an alternative argument, that $B_r$ can only grow through Tayler instability until the latter is quenched by magnetic tension \citep{Fuller19}. Further investigation is certainly needed to clarify the complicated saturation mechanisms of Tayler instability \citep{ji2023}. One way or the other though, a sufficient requirement for the AM transport prediction Eq.~\ref{scaling} to hold is, rather than Eqs.~\ref{lr} and \ref{dB} be independently satisfied, that their combination be true:
\begin{equation}
\fb{B_r}{B_\phi} \sim \fb{B_\phi}{\sqrt{\rho\mu}rN}.
\label{combi}
\end{equation}
(Indeed, replacing Eq.\ref{combi} in the induction equation dominant balance Eq.\ref{ind} is sufficient to derive Eq.\ref{scaling}.) The scaling law Eq.\ref{combi} turns out to be satisfied in our turbulent Tayler-Spruit dynamo simulations: Fig.\ref{fig_divB}(upper panel) shows the ratios $B_r/B_\phi$ against $B_\phi/\sqrt{\rho\mu}rN$, as crudely estimated from the time- and volume-averaged magnetic energy contained in the poloidal and toroidal axisymmetric magnetic field components, respectively. Even though both ratios  vary by only one order of magnitude throughout all the simulations, the scaling law Eq.\ref{combi} was found to hold over more than two decades in the magnetic field amplitude. It is rather more delicate to accurately assess the robustness of scaling Eq.\ref{lr} in our simulations, due to the difficulty of finding an objective (and automatic) criterion to precisely measure the width of the most-active dynamo region. Despite this caveat, the typical width is always found to be of order $0.1$ dimensionless units in our simulations (e.g. Fig.~\ref{cyl2sph}), which, given the unknown geometric prefactors in the various scaling laws, is roughly compatible with the magnitude of the ratio $B_\phi/(\sqrt{\rho\mu}rN)$.

Importantly, the Tayler-Spruit dynamo simulations 
presented here and in \cite{PetitdemangeMG23} are found in good agreement with \ref{scaling}, as illustrated in Fig.\ref{fig_divB} (bottom panel). To test the scaling law against our simulations, we measure the dimensionless magnetic torque
\begin{equation}
G_{mag}=\int_{{\cal S}(r)}\fb{sin\theta \, T(r)}{\rho\nu^2} \, d{\cal S},
\end{equation}
with ${\cal S}(r)$ the sphere of radius r centered at the origin, in the fluid region of the most intense magnetic activity and compare it with the scaling law
\begin{equation}
G_{mag} \sim {\cal N} \equiv r^{\frac{5}{2}}  \fb{(u_\phi\Omega)^{\frac{3}{2}}}{\rho \nu^2},
\label{Gmag}
\end{equation}
where $u_\phi$ is the axisymmetrix azimuthal velocity at radius $r$. (The numerical procedure used to calculate $G_{mag}$ from our simulations is detailed in the Supplementary Material of \citep{PetitdemangeMG23}.) The equivalence between \ref{Gmag} and Spruit's prediction \ref{scaling} is made possible because we use the following estimate of the dimensionless shear rate $q$:
\begin{equation}
q = \fb{r \pd_r \Omega}{\Omega}  \sim \fb{u_\phi}{\Omega l_r},
\label{qest}
\end{equation}
which we replace in \ref{scaling} to obtain
\begin{equation}
T  \sim \rho \Omega^2 r^{\fb{1}{2}} \fb{(u_\phi \Omega)^{\fb{3}{2}}}{N},
\label{scaling3}
\end{equation}
and thus \ref{Gmag}. One may wonder why $q$ is not constructed instead on the difference in rotation rates between the inner and outer sphere, i.e. the Rossby number:
\begin{equation}
q'=Ro=\Delta \Omega/\Omega.
\end{equation}
This is because our modeling choice of differentially rotating, rigid boundaries to prescribe a sheared velocity profile across the fluid domain, while numerically convenient, comes with a caveat: most of the shear is accomodated in the (very) thin, viscous boundary layer attached to the inner sphere. However, the peak of magnetic field intensity is met well away from the boundary, in a region where viscous effects are negligible. As a result, the shear that is effectively left across the dynamo region is significantly smaller than the total shear $q'$. It is almost completely accommodated through this region, for which the typical lengthscale of the Tayler instability \ref{lr} provides a natural width assessment (and one that is consistent with our simulations).

\begin{figure}
  \includegraphics[width=\linewidth]{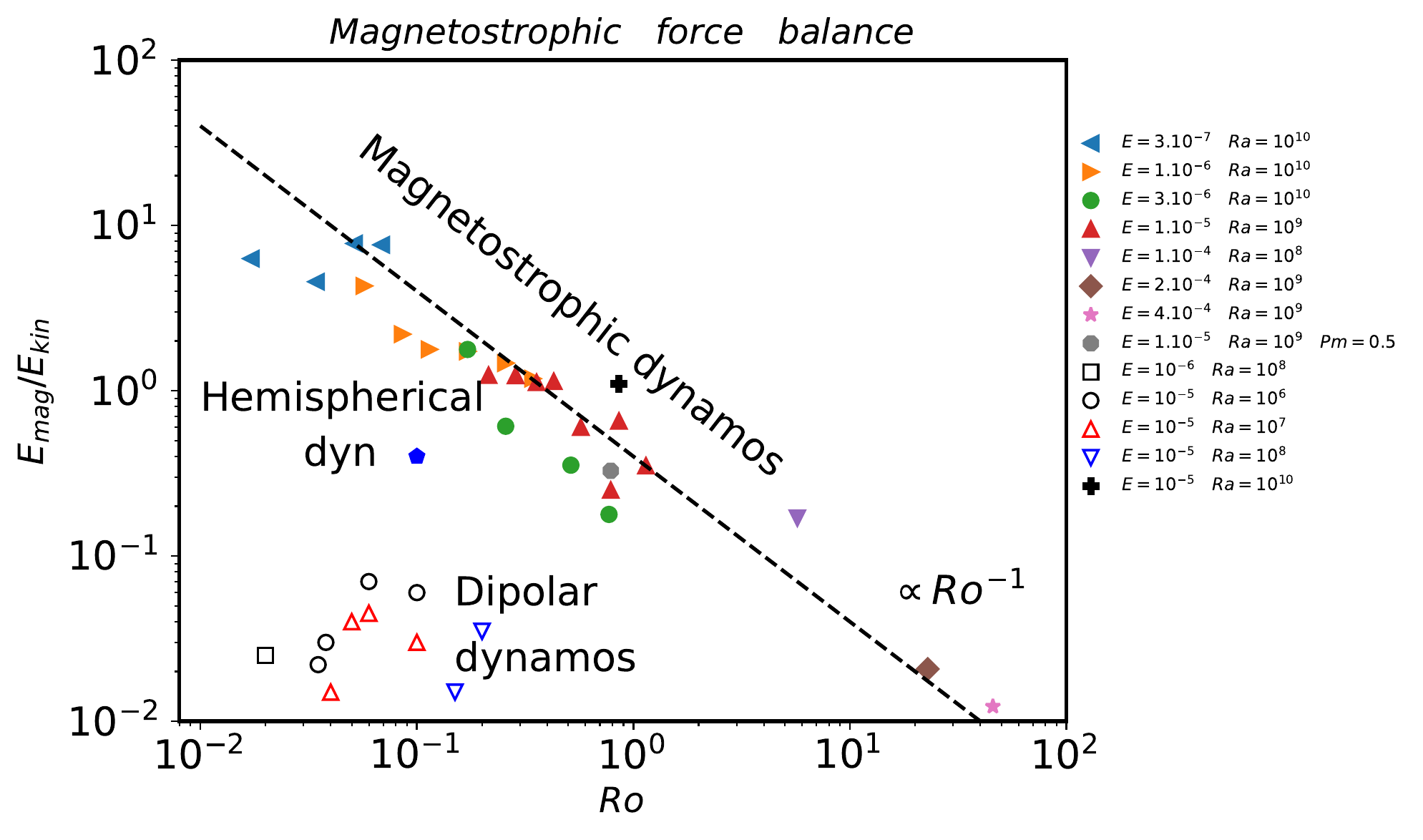}
\caption{Ratio of magnetic to kinetic energies as a function of the global shear parameter (dimensionless Rossby number), showing magnetostrophic scaling.}
\label{scalingMS}
\end{figure}

When flow motions are governed by magnetostrophic equilibrium, these considerations concerning the relevant shear estimates may be easily bypassed. The radial component of Navier-Stokes equation becomes at leading order, where Coriolis acceleration balances Lorentz force:
\begin{equation}
\rho \Omega u_\phi \sim \mu^{-1} B_\phi \fb{\partial B_\phi}{\partial r} \sim \fb{B_\phi^2}{\mu r}.
\label{MS}
\end{equation}
Combining \ref{MS} and \ref{combi} readily provide the alternative scaling \ref{scaling3} for the Maxwell stress:
\begin{align}
T &=\fb{B_r B_\phi}{\mu} = \left(\fb{B_r}{B_\phi}\right) \times \left(\fb{B_\phi^2}{\mu}\right) \sim \left(\fb{B_\phi}{\sqrt{\rho\mu}r}\right)(\rho \Omega u_\phi),\\
& \sim r^{\fb{1}{2}} \fb{(\Omega u_\phi)^\fb{3}{2}}{N}.
\end{align}
Note that in \ref{MS}, we have assumed that the dominant term in the Lorentz force is associated with the large-scale, axisymmetric (and slowly varying in radius) $B_phi$, rather than a fluctuation field $\delta B_\phi$ (which would vary over a scale $l_r$). While in our simulations, the former is indeed largely dominant, further investigation is required to determine whether the latter could affect governing balance at saturation in a (real) stellar interior. Our simulations turn out to fully satisfy magnetostrophic equilibrium: in Fig.~\ref{scalingMS}, the ratio between magnetic and kinetic energies obtained from our dataset is shown as a function of the Rossby number $Ro=\Delta \Omega/\Omega=Re\, E/\chi$. As proposed by data-inferred scaling laws put forward in different astrophysical contexts \citep{dormy16,AugustsonBT16,DormyOP18,KannaG19,RaynaudG20}, our strongly stratified Tayler-Spruit dynamos follow the magnetostrophic law where the ratio $E_{mag}/E_{kin}$ (calculated by averaging energies over the full fluid domain) is proportional to $Ro^{-1}$. This ratio is much smaller for the dipolar dynamos obtained with weaker stratification. Hemispherical dynamos also follow the strong-field scaling law when the ratio $E_{mag}/E_{kin}$ is only calculated in the hemisphere where dynamo action takes place. Note that the magnetostrophic scaling holds as the Rossby number varies over more than one decade and the ratio $E_{mag}/E_{kin}$ over two decades. For the lowest value of $Ro$ considered in our study, the magnetic energy exceeds the kinetic energy by more than one order of magnitude. Such a situation is reminiscent of the observations made for convection-driven dynamo simulations at low Ekman number \citep{schaefferJNF17,RaynaudG20}.

Finally, it is important to stress that, although our simulations use a finite thermal diffusivity $\kappa \gg \eta$, the magnetic torque is still found to scale like Spruit's diffusionless prescription \ref{scaling}. In other words, the transport of angular momentum achieved by Tayler-Spruit dynamos corresponds to fully developed MHD turbulence and does no longer depend on the fluid's molecular properties (kinematic, thermal or ohmic diffusivities) \citep{PetitdemangeMG23, Daniel2023}. 

\section{Conclusions}

Dynamo action in simplified models of radiative stellar layers can exhibit a rich diversity of magnetic fields amplitudes and morphologies, including hemispherical dynamos. In particular, the dynamos found in the (sufficiently) strongly-stratified regime are governed by magnetostrophic turbulence and achieve efficient AM transport, with measured Maxwell stresses in agreement with the diffusionless predictive scaling law of \cite{Spruit02}. The existence of a dynamo similar to the Tayler-Spruit mechanism is observed in the computationally challenging regime of Rayleigh number up to $Ra=10^{10}$, confirming that the results reported in \cite{PetitdemangeMG23} do pertain to even stronger stratification. Finally, intermediate scaling laws arising in the derivation of \cite{Spruit02} are carefully tested against our simulations results to assess the validity of the heuristic arguments in estimating magnetic stresses at saturation.

{It should be pointed out, however, that our simulations provide a picture that is relatively far from the idealised situations considered in previous theoretical studies. The fundamental role of turbulent fluctuations or the presence of complex magnetic field geometry considerably complicates the interpretation within the framework of these theories. Although the results reported here reproduce many of the predictions of the Tayler-Spruit dynamo, we believe that our turbulent simulations describe a more general scenario, relevant to stellar interiors, but for which the canonical Tayler-Spruit or AMRI descriptions are only asymptotic limiting cases.}

Full exploration of the parameter space is beyond the scope of the present study: the simulations presented here represent several millions of CPU hours, partly due to the necessity to integrate MHD equations over long times to rule out transient states, partly due to the computationally demanding flow regimes. Nevertheless, our numerical results already suggest that Tayler-Spruit dynamos pertain to a wider parameter regime as expected from the original theory: in particular, the analysis in \cite{Spruit02} relies on the hypothesis that $N\gg\Omega$, so that shellular approximation applies. The fact that our simulations with ratios $Q$ of order unity or below still agree with Spruit's original prescription is rather unexpected and shows the robustness of Tayler-Spruit mechanism for rapid rotators. Whether this scaling law is modified or not at extreme values of the Reynolds, Rayleigh, or Ekman numbers, or for different ordering in the diffusivity ratios, remains however to be discovered, and it would be certainly desirable to continue the numerical exploration of the parameter space as far as possible toward realistic parameter values. 

\begin{acknowledgements}
The authors wish to thank Sacha Brun, Jim Fuller, Daniel Lecoanet, Frank Stefani and Miguel-Angel Aloy Toras for useful discussions. This work was granted access to the HPC resources of MesoPSL funded by the Region Ile-de-France and the project Equip@Meso (reference ANR-10-EQPX-29-01) of the programme Investissements d'Avenir supervised by the Agence Nationale pour la Recherche. LP acknowledges financial support from ``Programme National de Physique Stellaire'' (PNPS) of CNRS/INSU, France. FM acknowledges financial support from the French program `T-ERC' managed by Agence Nationale de la Recherche (Grant ANR-19-ERC7-0008-01). CG acknowledges financial support from the French program `JCJC' managed by Agence Nationale de la Recherche (Grant ANR-19-CE30-0025-01).

\end{acknowledgements}

\bibliographystyle{aa} 
\bibliography{ref} 

\end{document}